\documentclass[aps,prx,a4paper,showpacs,amsmath,amssymb,superscriptaddress,floatfix,footinbib,twocolumn]{revtex4-2}

\usepackage{hyperref}
\hypersetup{colorlinks,linkcolor=blue,urlcolor=blue,citecolor=blue}

\usepackage{bbm}

\usepackage[normalem]{ulem}
\usepackage{tabularx}
\usepackage{graphics,graphicx}
\usepackage{epsfig}

\usepackage{xcolor}
\usepackage{float}
\usepackage{adjustbox}

\usepackage{dsfont}

\usepackage{amsmath,amssymb,mathrsfs}
\usepackage{qcircuit}
\usepackage{verbatim}

\usepackage{graphicx} % Required for inserting images

\newcommand{\code}[1]{{[[#1]]}}

\begin{document}

\title{Scaling the Automated Discovery of Quantum Circuits \\ 
       via Reinforcement Learning with Gadgets}

\author{Jan Olle}
\affiliation{Max Planck Institute for the Science of Light, Erlangen, Germany}

\author{Oleg M. Yevtushenko}
% \affiliation{Max Planck Institute for the Science of Light, Erlangen, Germany}
\affiliation{Friedrich-Alexander-Universit{\"a}t Erlangen-N{\"u}rnberg, Germany}

\author{Florian Marquardt}
\affiliation{Max Planck Institute for the Science of Light, Erlangen, Germany}
\affiliation{Friedrich-Alexander-Universit{\"a}t Erlangen-N{\"u}rnberg, Germany}

\begin{abstract}
Reinforcement Learning (RL) has established itself as a powerful tool for 
designing quantum circuits, which are essential for processing quantum 
information. RL applications have typically 
focused on circuits of small to intermediate complexity, as computation 
times tend to increase exponentially with growing circuit complexity. This 
computational explosion severely limits the scalability of RL and casts 
significant doubt on its broader applicability.
In this paper, we propose a principled approach based on the systematic discovery and 
introduction of composite gates -- {\it gadgets}, that enables RL scalability, 
thereby expanding its potential applications. As a case study, we explore the 
discovery of Clifford encoders for Quantum Error Correction. We demonstrate 
that incorporating gadgets in the form of composite Clifford gates, in addition to standard CNOT and 
Hadamard gates, significantly enhances the efficiency of RL agents. Specifically, 
the computation speed increases (by one or even two orders of magnitude), enabling RL to 
discover highly complex quantum codes without previous knowledge.
We illustrate this advancement with examples of QEC code discovery with parameters $ [[n,1,d]] $ for 
$ d \leq 7 $ and $ [[n,k,6]] $ for $ k \leq 7 $. We note that the most complicated 
circuits of these classes were not previously found. We highlight the advantages 
and limitations of the gadget-based approach. Our method paves the way for scaling the RL-based automatic discovery of complicated quantum circuits for various tasks, which may
include designing logical operations between logical qubits or discovering quantum algorithms.
\end{abstract}

\date{\today}

\maketitle

\begin{figure*}
    \centering
    \includegraphics[width=0.9\linewidth]{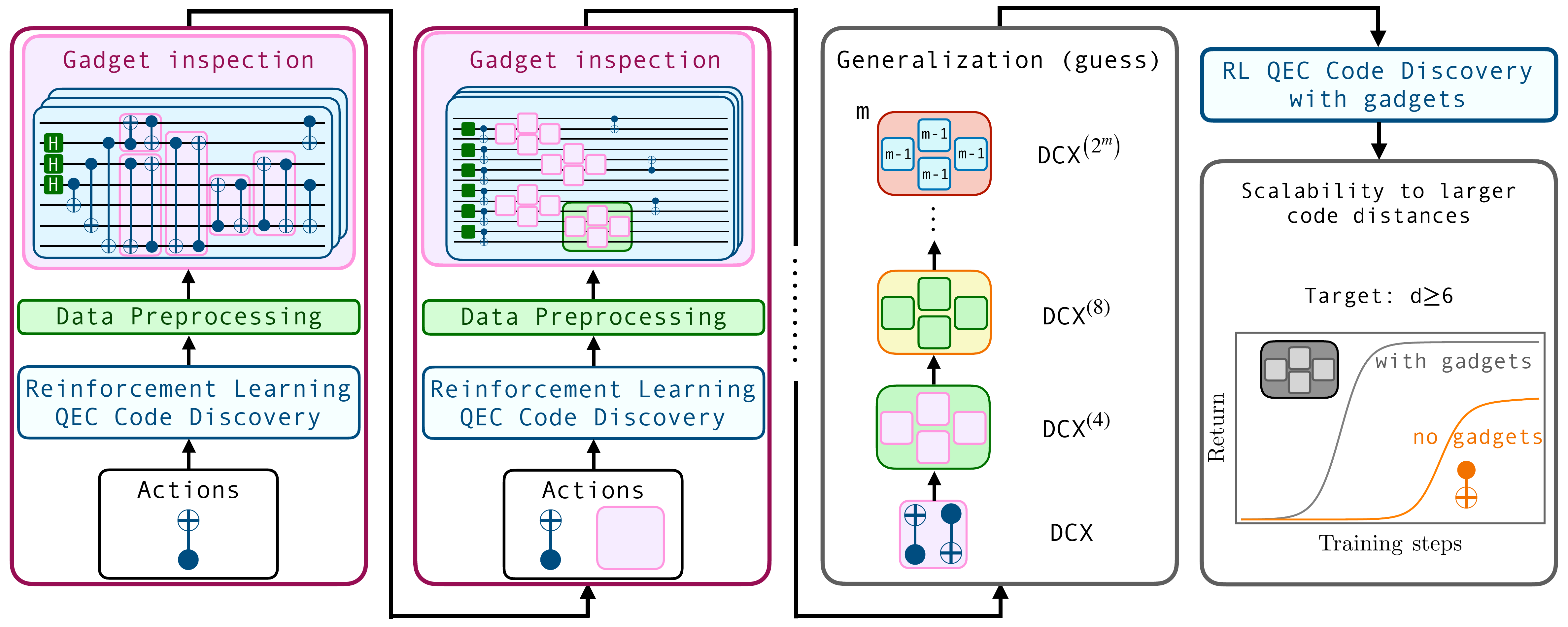}
    \caption{\textbf{Conceptual scheme of our workflow.} The process begins with basic CNOT gates as actions for the reinforcement learning (RL) agent. After finding many encoding circuits, these are preprocessed (explained in more detail in Figure.~\ref{fig:conceptual_fig_RL_preprocessing}) and further inspected in order to visually identify recurring patterns, which we call gadgets. These gadgets are then given as further actions to a new generation of RL agents, and the process gets repeated. Eventually, we notice a recursive pattern and identify a rule to generalize gadget construction. Finally, thanks to these powerful gadgets, we are able to scale the RL strategy to discover QEC codes with larger code distances than what we are able to find when only single CNOTs are available as actions.}
    \label{fig:conceptual_fig_workflow}
\end{figure*}

\begin{figure*}
    \centering
    \includegraphics[width=0.9\linewidth]{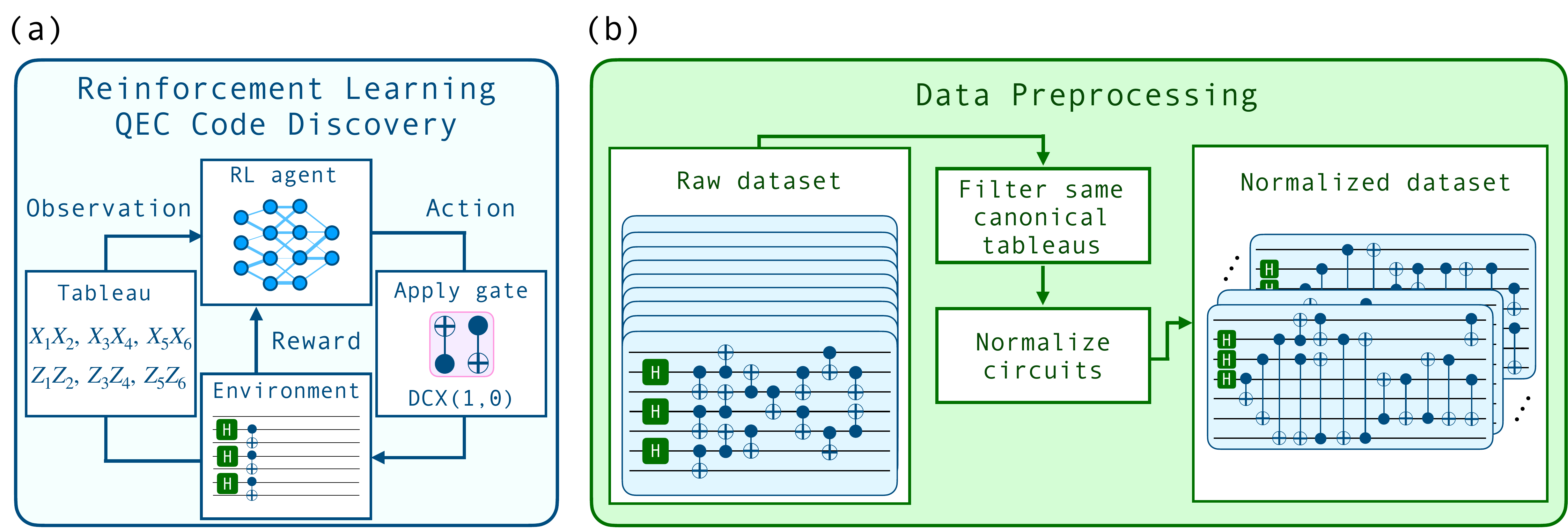}
    \caption{\textbf{Scheme of the reinforcement learning and data preprocessing modules.} (a) QEC code and encoder discovery with reinforcement learning. The RL agent's task is to build a circuit that is able to correct a target list of errors, which enter through the reward. To train the agent, we use PPO, which is an actor-critic method with two neural networks. They both receive as observation a binary representation of the tableau of the circuit at that given point in time. The actions are discrete and correspond to applying either a single CNOT gate or more complex gates built from multiple CNOTs, whose possible control and target qubits are determined by the available qubit connectivity. (b) The data preprocessing module. It consists of two steps: a filtering step where we only store one circuit per unique canonical tableau, and a normalization step where the remaining circuits are brought to a normalized form (for more details see the main text). This procedure is crucial to remove redundancy and complexity from our dataset and enables a more tractable visual gadget recognition process.}
    \label{fig:conceptual_fig_RL_preprocessing}
\end{figure*}

\section{Introduction}

As the field of quantum technology matures, the number of qubits that have to be controlled and orchestrated is growing~\cite{mohseni2025buildquantumsupercomputerscaling,google2024quantum,aghaee2025scaling}. To extract useful value from them, quantum circuits have to be designed with different specific purposes. This essential task, commonly known as quantum circuit synthesis, consists of the following: given a target abstract quantum operation (which can be either preparing a quantum state,  preparing a desired unitary operation, or implementing an entire quantum algorithm), the task is to find a sequence of gates that can run natively in the quantum processor and that produce the desired quantum operation with high success rate.

Techniques from machine learning have been identified as powerful tools to tackle the quantum circuit synthesis task~\citep{PhysRevX.8.031084, doi:10.1073/pnas.1714936115, krenn2023artificial, Arrazola_2019, he2023gnn, Zhang_2021, foesel2021quantumcircuitoptimizationdeep, moro2021quantum, PhysRevLett.125.170501, Sarra_2024, Preti_2024, kundu2024easyhardtacklingquantum, Furrutter_2024, ruiz2024quantumcircuitoptimizationalphatensor}. From these, reinforcement learning (RL)~\cite{sutton2018reinforcement} is particularly well suited for solving sequential decision-making tasks where an optimal solution is a priori not known. However, as the complexity of the task increases (typically scaling either the number of qubits or the circuit size), RL faces serious challenges and can even eventually fail. 

Two strategies have been explored in order to ease the scalability of RL approaches for the quantum circuit synthesis task. The first one is curriculum RL~\cite{ostaszewski2021reinforcementlearningoptimizationvariational,patel2024curriculumreinforcementlearningquantum}. There, an agent learns to solve the complex task by gradually solving tasks of increasing difficulty, transferring the knowledge throughout the different stages. An alternative strategy is to identify frequent and compact subroutines used by the agent, which are commonly referred to as \textit{gadgets} (a term originating in complexity theory in computer science). In \cite{Trenkwalder_2023}, gadgets from an RL agent's strategy for producing quantum entangled states are identified, yet the goal was the interpretability of those actions. In \cite{ruiz2024quantumcircuitoptimizationalphatensor}, predetermined gadgets were given to RL agents to optimize quantum circuits, but no new gadgets were discovered. More recently, the iterative discovery and use of gadgets for quantum ground state preparation has been developed~\cite{kundu2024easyhardtacklingquantum} following developments in program synthesis for quantum circuit synthesis~\cite{Sarra_2024}. A different, yet related approach is called projective simulation~\cite{briegel2012projective}, an alternative version of RL, where an agent simulates potential future scenarios through random walks in a network of memory clips before taking action and where new clips can be generated. This has been applied, for example, to long-distance quantum communication problems~\cite{PRXQuantum.1.010301}.

While the general idea of employing gadgets in itself is not novel, there is no unique way of systematically constructing and using them. In the quantum domain, the iterative use of discovered gadgets to solve more complex tasks has only been mildly successful – only scaled to a handful of qubits~\cite{kundu2024easyhardtacklingquantum}. In contrast, we will present a concrete RL-based implementation of this idea that is able to scale to multiple dozens of qubits. The chosen domain of application, both due to its importance and its viability, is the automated discovery of encoding circuits for Quantum Error Correction (QEC).

With the ongoing race towards showing scalable experimental demonstrations of QEC~\cite{krinner2022realizing,ryan2021realization,postler2022demonstration,cong2022hardware,GoogleQuantum2023,sivak2023real,google2024quantum}, it is of particular interest to design QEC protocols - which can also be understood as a quantum circuit synthesis task - with scalable algorithmic methods. Operating as a virtual scientist, an RL agent can systematically explore the space of possible codes through trial and error, with the advantage of allowing the human to have precise control over which structural constraints to enforce or relax. This data-driven approach enables the investigation of more creative QEC strategies that might elude conventional analytical methods.

Recent advances have demonstrated the viability of RL for QEC tasks \cite{PhysRevX.8.031084,Nautrup_2019,mauron2023optimization,su2023discovery,QuantumLego,andreasson2019quantum,sweke2020reinforcement,colomer2020reinforcement,fitzek2020deep,metz2023self,olle_2024,zen2024quantum,eickbusch2024demonstratingdynamicsurfacecodes,freire2025optimizinghypergraphproductcodes, Puviani_2025}. Specifically, in the important domain of stabilizer codes, recent promising state-of-the-art results have illustrated the discovery of codes and their encoding circuits up to distance 5~\cite{olle_2024}. However, scaling these approaches to more complex scenarios has proved challenging, suggesting that vanilla RL strategies may not be sufficient for discovering higher-distance codes.

In this work, we present a scalable approach to QEC code discovery using RL with gadgets: frequently occurring subroutines that can be abstracted as single actions for a higher-level RL agent. By identifying and leveraging these computational motifs, we demonstrate the successful discovery of distance-7 codes from scratch, marking a significant advancement in automated QEC code design. This achievement suggests a promising path toward scaling the discovery of QEC codes to the distances required for practical quantum computing applications with RL.

This paper is organized as follows: in Section \ref{sect:background} we provide the theoretical background for stabilizer and CSS codes and reinforcement learning. In Section \ref{sect:method} we describe our approach to discover and use gadgets in the task of QEC encoder discovery. Our results are presented in Section \ref{sect:results}, and we provide a discussion and conclusions in Section \ref{sect:discussion}.

\section{Background}
\label{sect:background}

\subsection{Stabilizer and CSS codes}

The stabilizer formalism~\cite{gottesman1997stabilizer} provides a resource-efficient description of quantum states which is particularly useful for QEC. 
The central idea is to describe a quantum state by listing the set of operators that \textit{stabilize} it, i.e. of which that state is an eigenvector with eigenvalue $+1$. 
When working with qubits, the useful set of operators to consider are Pauli strings: Kronecker products of the Pauli matrices $I,X,Y,Z$ over all qubits.
With this choice, given $n$ qubits, a quantum state can be described by listing $n$ Pauli strings. 
Importantly, Pauli strings can be represented as binary arrays of size $2n$~\cite{Aaronson_2004}, meaning that $2n^2$ bits suffice to represent a quantum state.
The \textit{weight} of a Pauli string is its non-trivial support in the space of qubits.

The stabilizer formalism also allows to efficiently describe the time evolution of stabilizer states. However, it must be restricted to unitaries that map Pauli strings to Pauli strings in the Heisenberg picture. By definition, these Pauli-preserving unitaries are called Clifford gates and can be generated by the Hadamard \(H\), the Phase \(S\) and the CNOT gates~\cite{Aaronson_2004}.

The stabilizer formalism can also describe subspaces (called codes) and their time evolution. A code that encodes \(k\) logical qubits into \(n\) physical qubits is a \(2^k\)-dimensional subspace (the \textit{code space} \(\mathcal{C}\)) of the full $2^n$-dimensional Hilbert space. It is completely specified by a set $\{g_i\}$ of $n-k$ Pauli strings that \textit{stabilize} it. In fact, these $n-k$ Pauli strings generate a group denoted by $S_\mathcal{C} = \langle g_1, g_2, \dots, g_{n-k} \rangle$, which is called the stabilizer group of $\mathcal{C}$. In order to describe such codes, one needs $2 n (n-k)$ bits.

Quantum codes are classified according to how many errors they can detect/correct, according to the Knill-Laflamme conditions~\cite{BDW96,PhysRevA.55.900}. The standard classification is constructed by decomposing arbitrary errors into Pauli strings and checking the smallest weight that cannot be detected. Explicitly, a quantum code that can detect all Pauli strings of up to weight $d-1$ but that fails to detect at least one Pauli string of weight $d$ is called a distance $d$ code. This results in a code of distance $d$ being able to correct all errors up to a weight $t$ such that $d = 2t + 1$ \cite{gottesman1997stabilizer}. We follow standard notation and denote quantum codes of distance \(d\) that encode \(k\) logical qubits into \(n\) physical qubits as \(\code{n,k,d}\).

CSS codes~\cite{Steane1996,Shor1996}, named after A. R. Calderbank, P. Shor, and A. Steane, are a subclass of stabilizer codes with very useful properties. By definition, they are generated by Pauli strings containing either only $X$'s or only $Z$'s (apart from $I$). We refer to the $X$-type generators of a CSS code as $G_X$ and the $Z$-type ones as $G_Z$. For instance, Steane's $\code{7,1,3}$ code has generators $G_X = \{ IIIXXXX,~IXXIIXX,~XIXIXIX\}$ and $G_Z = \{ IIIZZZZ,~IZZIIZZ,~ZIZIZIZ\}$. Surface codes are also CSS codes.

By construction, CSS codes detect $X$-type and $Z$-type errors \textit{independently}. This implies that $Y$-type errors are identified when X and Z-type stabilizer measurements fire simultaneously. In practice, when evaluating a code's error correction capabilities, it suffices to verify which $X$-type and $Z$-type Pauli strings can be detected. For instance, a code of distance $d$ can detect all such Pauli strings of up to weight $d-1$. There are
\begin{equation}
    \text{num} \left( \left\{E_\mu\right\}\right) = \sum_{w=0}^{d-1} \binom{n}{w}~ \label{eq:numE_CSS}
\end{equation}
such Pauli strings of $X$-type and an equal number of $Z$-type.

\subsection{Reinforcement Learning}

Reinforcement Learning (RL)~\cite{sutton1999policy, sutton2018reinforcement} provides a framework for identifying optimal action sequences in sequential decision-making tasks.
The task to solve is encoded by a scalar quantity called the \textit{reward} \(r\), which has to be carefully chosen to guide the algorithm towards the goal that we are interested in achieving.
The entity making these decisions is called the \textit{agent}, and in our work is realized with a neural network.
The agent interacts with an \textit{environment}, which is the physical system of interest or a simulation of it. 
In each time step \(t\), the environment’s state \(s_t\) is observed. 
Based on this observation, the agent takes an action \(a_t\) which affects the state of the environment and yields a reward signal $r_t$.
Common to all RL algorithms is the objective of maximizing the expected cumulative reward (called the return), \( \mathbb{E} \left[ \sum_t r_t \right]\) over a trajectory. 
A \textit{trajectory} is a sequence of state, action and reward triples that the agent experiences from an initial state ($t=0$) to a terminal state ($t=T$).

There are many different methods in RL to do this optimization. Some of the most successful ones are under the umbrella of policy gradient algorithms~\cite{sutton1999policy}. 
A policy is a function $\pi = \pi_\theta (a_t|s_t)$ that defines the strategy of the RL agent and that is mathematically defined as a probability distribution of choosing action \(a_t\) given observation \(s_t\), according to the neural network with parameters \(\theta\).
Policy gradient algorithms optimize the policy $\pi_\theta$ by maximizing the expected return with respect to the parameters \(\theta\) with gradient ascent. Within policy gradient methods, actor-critic algorithms~\cite{konda1999actor} are the most commonly used ones. The idea is to have two neural networks that are trained simultaneously: one for the policy (actor network), and a second one called the critic network that measures how good the action from the policy network was. In this paper, we use a state-of-the-art policy-gradient actor-critic method called Proximal Policy Optimization (PPO)~\cite{schulman2017proximal}, which is particularly well-suited for problems with discrete actions.

We closely follow the implementation of \cite{olle_2024} with some minor differences. The reward function is based on the Knill-Laflamme error correction conditions~\cite{BDW96,PhysRevA.55.900}, but we take the \textit{difference} in Knill-Laflamme conditions between two consecutive timesteps as an instantaneous reward instead of their current value. Explicitly, we define the Knill-Laflamme sum $\Sigma_{KL}$ as
\begin{equation}
    \Sigma_{KL} = \sum_\mu \lambda_\mu K_\mu~,
\end{equation}
where $\mu$ is an index that runs over the number of error operators that should be detected (see Eq.~\eqref{eq:numE_CSS}), $K_\mu$ is either 0 or 1 depending on whether the corresponding error operator $E_\mu$ can be detected (0), or not (1); and $\lambda_\mu$ are real positive hyperparameters weighing each corresponding error that for now can be thought to be their likelihood $p_\mu$. By definition, $\Sigma_{KL}$ is either positive or null. In the former case, some errors are undetectable and only when $ \Sigma_{KL} $ is zero can we guarantee that all errors $\{E_\mu\}$ satisfy the Knill-Laflamme conditions and, hence,
can be detected. In this work, we use the instantaneous reward
\begin{equation}
    r_t = - \left[\Sigma_{KL}(t) - \Sigma_{KL}(t-1) \right]~, \label{eq:instantaneous_reward}
\end{equation}
which is positive if more errors are detected at the current timestep than at the previous one, and negative otherwise. 

The second implementation difference with respect to \cite{olle_2024} is a modification of the PPO algorithm itself according to \cite{naegele2024tacklingdecisionprocessesnoncumulative} (MAXPPO). 
In particular, this algorithm maximizes 
\begin{equation}
    \mathbb{E}_\pi \left[ \max_{k \in [0,T]} \sum_{t=0}^k r_t \right]~,
\end{equation}
with $r_t$ given by Eq.~\eqref{eq:instantaneous_reward}. The reason that we use this algorithm is that it was designed to find the RL state with the lowest cost found during a trajectory, i.e. a QEC code in our application. Intuitively, this algorithm allows the agent to explore more freely the possible space of solutions thanks to not receiving negative rewards when trying to escape from local minima.

All other implementation details, such as neural network architecture or hyperparameters used, are identical to those used in \cite{olle_2024}.

\section{Method: RL with gadgets}
\label{sect:method}
\subsection{Motivation}

A fundamental challenge in discovering quantum error correction codes is the exponential growth in the search space as we scale the target code parameters $\code{n,k,d}$.
In the context of RL-guided discovery of encoding circuits, it was argued in~\cite{olle_2024} that a \textit{region of opportunity} exists  for code parameters in the range of $ 25 \lesssim n \lesssim 60$ and $6 \leq d \leq 8$. However, when starting to probe this regime, we found naïve RL training runs for $n \gtrsim 25$, $d \geq 6$ to be completely unfruitful. We attribute this to two main factors:
\begin{enumerate}
    \item Hierarchy of error operators and likelihoods: the number of errors increases exponentially (see Eq.~\eqref{eq:numE_CSS}), but their likelihood decreases exponentially ($p^d$), giving very weak reward signals for errors with higher weight. 
    \item More complex encoding circuits are needed, meaning many more gates. This leads to the so-called long horizon problem in RL: the idea that the search space of trajectories grows as $(n_A)^T$ with the number of actions $n_A$ and trajectory length $T$.
\end{enumerate}

These issues seriously hamper the scalability of RL and cast doubts on its broader applicability for the automated discovery of quantum circuits in the more challenging situations of larger code parameters.

One could consider two strategies going forward. The first one would be to change the reward function to take into account the hierarchy of the different error operators that participate in the QEC conditions. This would alleviate problem 1, but would not make the large horizon problem any easier. We found no obvious way to design such a reward function and leave this as an interesting area of future research. The second strategy consists in allowing the agent to use more complex actions. This has the obvious benefit of taming the large horizon problem by needing smaller trajectories to solve the problem, but in principle does not alleviate the error hierarchy problem affecting reward signals. 

In this work, we describe a successful implementation of the second strategy which allows one to overcome both issues  mentioned above. We provide a conceptual illustration of the entire procedure in Fig.~\ref{fig:conceptual_fig_workflow}. It consists of the repeated application of a computational block containing an automated search of encoding circuits with RL and a further processing of those circuits that helps identify motifs. These motifs, which we call gadgets, are reused as actions of the next generation of RL agents and new gadgets are found. This whole process is iterated until we find a rule that allows us to generalize the construction of gadgets. Thanks to these, we are able to scale the RL strategy to discover QEC codes with code distances larger than those possible when primitive Clifford gates were used.

\subsection{Building gadgets}

To enable the scalability of RL to discover more complex circuits, we have expanded the set of allowed actions which are provided to the RL agent. In the vanilla version~\cite{olle_2024}, the agent has access to primitive Clifford gates such as the Hadamard or the CNOT gate. The strategy is to build new actions as ''composite Clifford gates'' consisting of several of the primitive Clifford gates, see Fig.\ref{fig:conceptual_fig_RL_preprocessing}. We refer to the composite gates as {\it gadgets}.

Building gadgets is a complex task due to their inherent combinatorial nature. For instance, given $n$ qubits, all-to-all connectivity and two CNOTs, there are $\binom{n}{4}$ possible gadget configurations (corresponding to the 4 positions where the controls and targets can be placed). From all of these, it is not clear at all which are useful gadgets and which are not. In addition, since CNOTs acting on different qubit subsets commute, many of these would be equivalent. Thus, we would ideally want to build useful gadgets whose components – the individual Clifford gates – do not commute along the quantum circuit, i.e. along the ``time-axis''. We refer to such gadgets as ``static''.

We do not know {\it a priori} an optimal way of constructing useful gadgets.
We thus start by analyzing the automatically discovered encoders for small and medium-size codes, e.g. the codes with $ 3 \leq d \leq 5 $ and $k = 1$, discovered by RL \textit{without} gadgets. Such an analysis consists of several steps. 

\subsubsection{Pre-processing simple discovered circuits and identifying the simplest gadgets}

\begin{figure}
    \centering
    \includegraphics[width=0.95\linewidth]{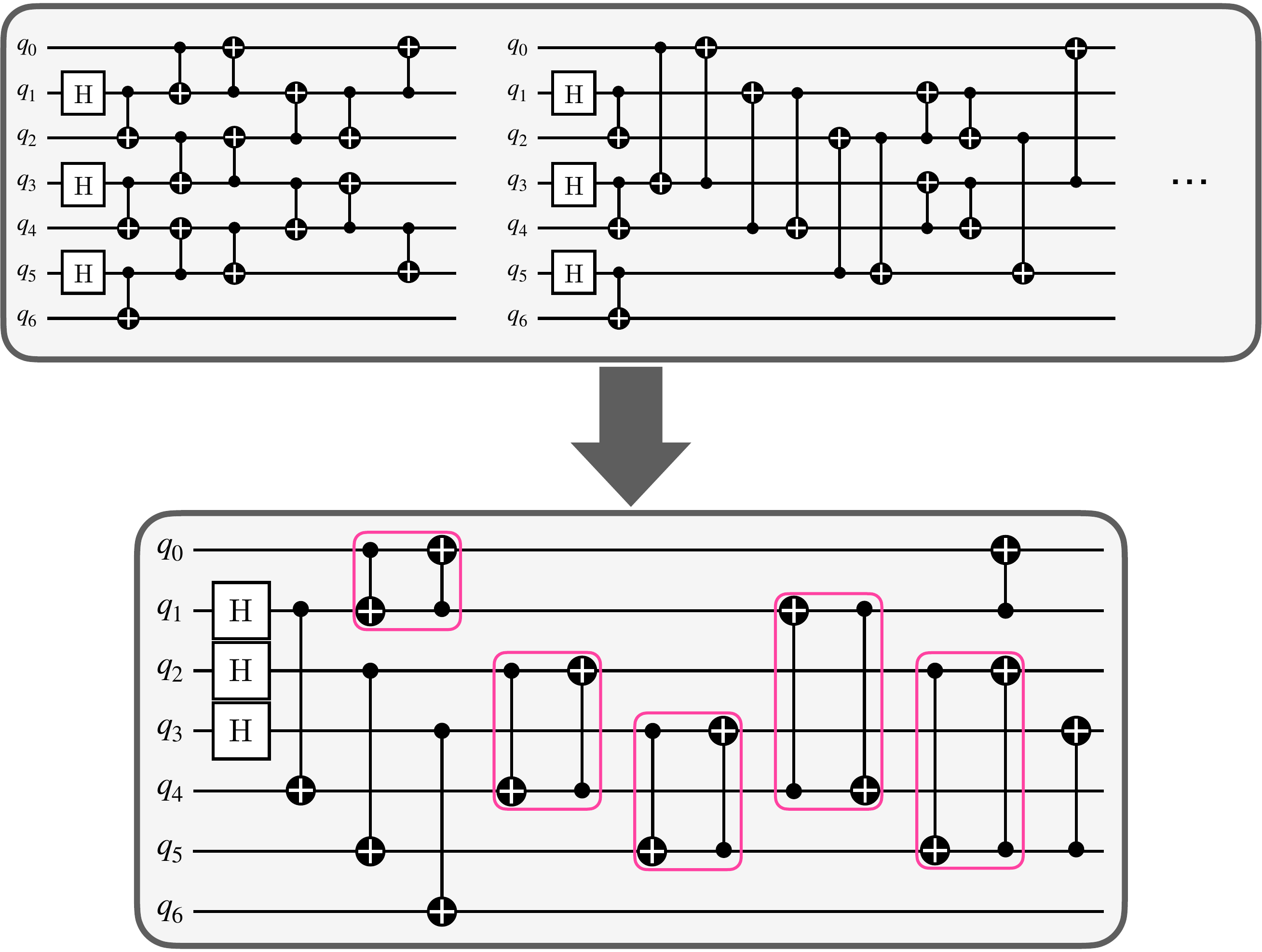}
    \caption{Example of a DCX gadget pattern in a [[7,1,3]] code. First we illustrate the second step in the data preprocessing pipeline where two circuits with different canonical tableaux get mapped to the same circuit in normalized form. After preprocessing the raw dataset into circuits in normal form, one remains with a handful of circuits that are representatives of all patterns found by the RL agent. Here, we highlight DCX patterns.}
    \label{fig:motifs}
\end{figure}

First, we start with $d=3$ and $n=7$ and $n=9$ ($k=1$). For these code parameters, we launch $O(100)$ training runs to extract $O(100)$ raw encoding circuits. Some of these will constitute equivalent codes, in the sense that they will correspond to the same canonical tableau - a standardized $(n-k)\times 2n$ binary matrix representation of the code's stabilizer generators where each row encodes a different generator. Using this representation is useful because two codes are equivalent if their canonical tableaux are identical up to qubit permutations. 
We thus filter this raw dataset by keeping only a single circuit per canonical tableau instance and discard the rest. 
At this moment, we have a dataset of inequivalent circuits up to qubit permutations, cf. Fig.\ref{fig:motifs}.

The next step is to reduce the qubit permutation multiplicity of the dataset by bringing the circuits to a standardized (or normal) form as follows. First, label the logical qubit to be $0$ and the qubits where Hadamard gates are placed to be $1,2,\dots,\text{num}_H$. Then, temporally traverse the circuit and relabel the participating qubits in the gates. If there is a qubit that has not yet been relabeled, assign the next available label starting from $\text{num}_H + 1$. The result is a normalized dataset of inequivalent circuits (see Fig.\ref{fig:conceptual_fig_RL_preprocessing}) that is much smaller than the raw dataset but that essentially contains all its information. Crucially, this enables further visual inspection by a human in order to detect repeating patterns that will constitute the gadgets. 

Next, we visually identify recurring structural patterns of gates in these normalized circuits. An example of a motif which occurs in the $\code{9,1,3}$ code is shown in Fig.\ref{fig:motifs}. The simplest repeating object is a combination of two CNOT gates: the double-CNOT gate (DCX), see Fig.\ref{fig:gadgets}. Note that this gadget is static in the sense that its individual constituents (the CNOTs) do not commute. Moreover, DCX gadgets can have two orientations, 
see Fig.\ref{fig:gadgets}. Thus, given the underlying connectivity graph of CNOTs, there are as many DCX gadgets as CNOTs. This is important when we allow DCX gadgets as new actions, as the number of new actions is much smaller than the naïve number of possible gadgets obtained from a combinatorial argument.

\subsubsection{Simple gadgets as building blocks for \\
               next-generation gadgets}

Having identified the simplest gadgets, we now allow new RL training runs to contain DCXs as possible actions, together with individual CNOTs. Launching these runs in the regime of small code parameters ($n=7,9$, $k=1$, $d=3$) does not lead to interesting insights due to the rather simple structure of the encoding circuits. We thus increase to $n=13$ and $d=4$, keeping $k=1$ fixed, and we gather $O(100)$ automatically discovered circuits with DCX gadgets (second column of Fig.\ref{fig:conceptual_fig_workflow}). After having pre-processed the dataset of circuits into its normalized form, we notice the frequent appearance of a pattern which consists of four consecutive DCX gates affecting neighborhoods of four qubits, see Fig.\ref{fig:gadgets}. 

We call these gadgets $\text{DCX}^{(4)}$ because they are built from DCX gates and affect 4 qubits. Interestingly, these new gadgets can also only have two different orientations and are static. Thus, the total number of $\text{DCX}^{(4)}$ gadgets, that could be added as new actions given the underlying CNOT graph connectivity, is upper-bounded by the number of different CNOTs. More precisely, it is given by the number of connected subgraphs of 4 qubits.

\subsubsection{Using powerful gadgets to scale the automated discovery of complex encoders with RL}

Motivated by the unexpected finding of an emergent descendant hierarchy of gadgets (from CNOT to DCX, to DCX$^{(4)}$), 
we have surmised that the nesting could develop further in the larger codes with $ d > 5 $ and $ k \gg 1 $.

\begin{figure}
    \centering
    \includegraphics[width=0.95\linewidth]{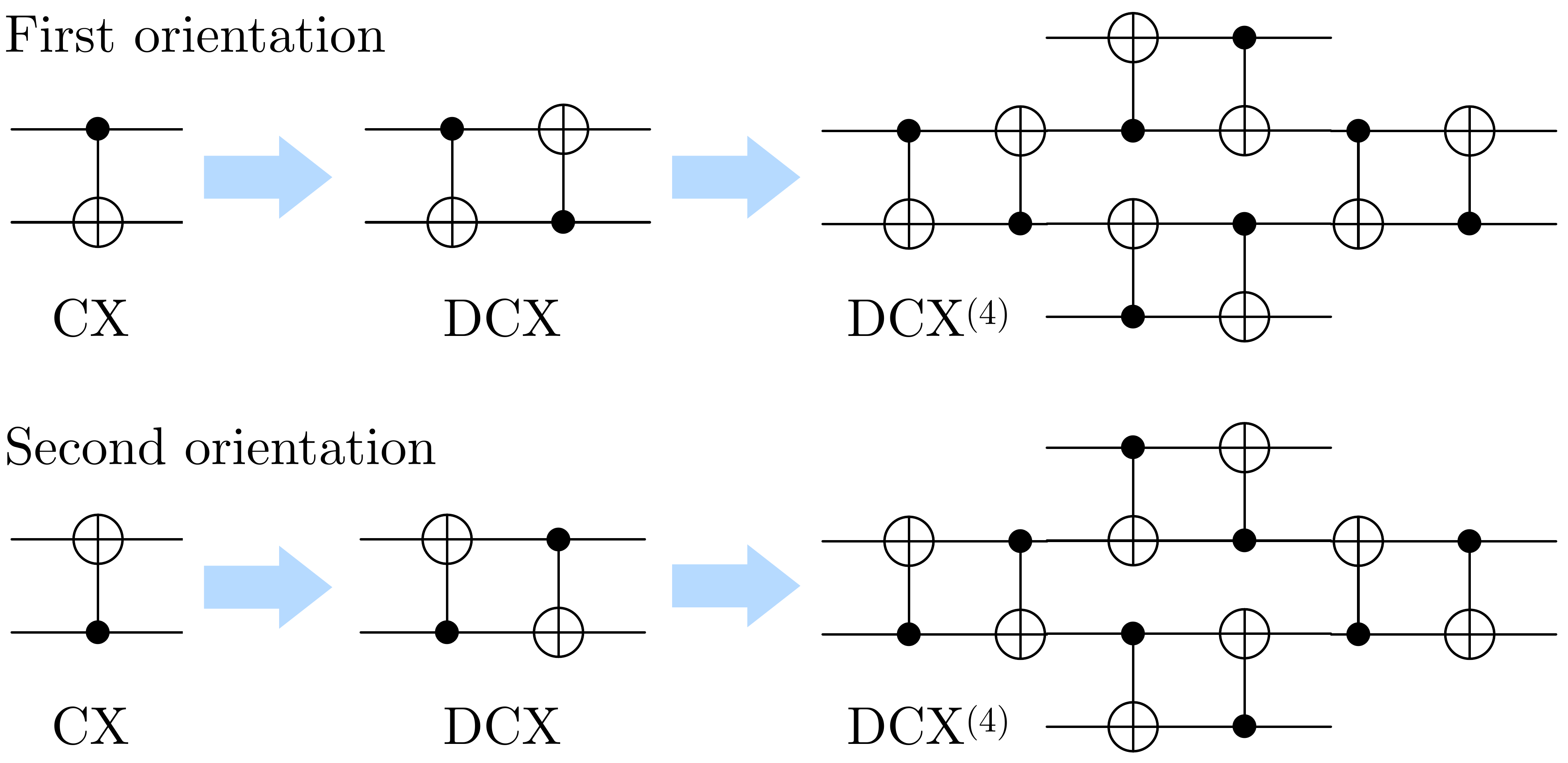}
    \caption{Examples of hierarchy of multiqubit gates which are provided to the RL agents as allowed actions. Note that the gadgets of the next generation involve more qubits. For example, DCX$^{(4)}$ uses four qubits.
    Depending on the position of control- and target qubits in the parent CNOT gate, one can obtain two different orientations of the descendant gadgets.}
    \label{fig:gadgets}
\end{figure}

Here, we conjecture that there is a tower of new descendant gadgets (DCX$^{(8)}$, DCX$^{(16)}$, etc.) constructed by analogy to how DCX$^{(4)}$ is built from DCX. Explicitly, we conjecture that the rule for gadget DCX$^{(2 m)}$ is constructed from four DCX$^{(2 [m-1])}$ assembled following the cross-pattern seen in Fig.\ref{fig:gadgets}. We also assume that this rule is the same for all gadget generations. This rule maintains the two properties of useful gadgets as actions in RL, namely: the fact that only two orientations are allowed and that they are static. Crucially, these keep the otherwise natural combinatorial proliferation of possible gadgets under control.

In the next Section, we demonstrate how our approach enhances the efficiency of RL agents and opens the avenue for its further scalability. We will also show how the codes found by the gadget-based RL compare to other well-known codes, such as surface codes and low-density parity-check codes, see Sect.\ref{sect:n_d-diagr} and \ref{sect:n_k-diagr} below.

\subsection{Details of implementation}

Several implementation techniques were crucial for the success of the scaling of the RL strategy with gadgets. Here, we detail the key elements that enabled efficient training and improved results.

To facilitate the discovery of QEC codes with larger distances, we implemented an automatic transfer learning between scenarios with increasing target code distance. This allows the agent to leverage patterns learned from simpler codes when constructing more complex ones. For instance, when the goal is to discover a $\code{n,k,d=6}$ code, the agent's first target is to find a $\code{n,k,d=4}$ code. After a few training epochs, the same agent is then requested to find a $\code{n,k,d=5}$ code, transferring the learned parameters from the $d=4$ task. Finally, after a few training epochs at $d=5$, the agent is tasked with finding the $\code{n,k,d=6}$ code that we were after. We have found this strategy to improve the efficiency and stability of training runs with respect to a cold start at $d=6$. This strategy can also be viewed as a form of curriculum learning, where we leverage our understanding of smaller quantum codes to facilitate the discovery of larger ones. 

The second implementation detail that we found to be very helpful - particularly when searching for codes with $k>1$ - consists of two ingredients: (i) place the logical qubit indices equally spaced,  alternating the Hadamard placing in the remaining, and (ii) use periodic boundary conditions in the qubit connectivity graph. These two design choices provide a more uniform configuration for the different qubits of the system. In addition, we also initialize the circuit with Bell pairs between adjacent qubits for qubits that are not placed in a logical index.

One scheme that we explored but found to not be crucial was weighing the reward with an additional term depending on which kind of gate/action was used. In particular, this strategy consisted of penalizing the agent when using too many gadgets. This stems from observations with training runs on simple codes where the tendency was for the RL agent to use a few gadgets at the beginning and simple CX gates afterwards. To encourage this behavior in agents for more complex encoders, we tried introducing a penalty when a gadget was used after a certain threshold timestep. Since we saw comparable performance to not using it, we decided to drop it.

\begin{figure}
    \centering
    \includegraphics[width=0.95\linewidth]{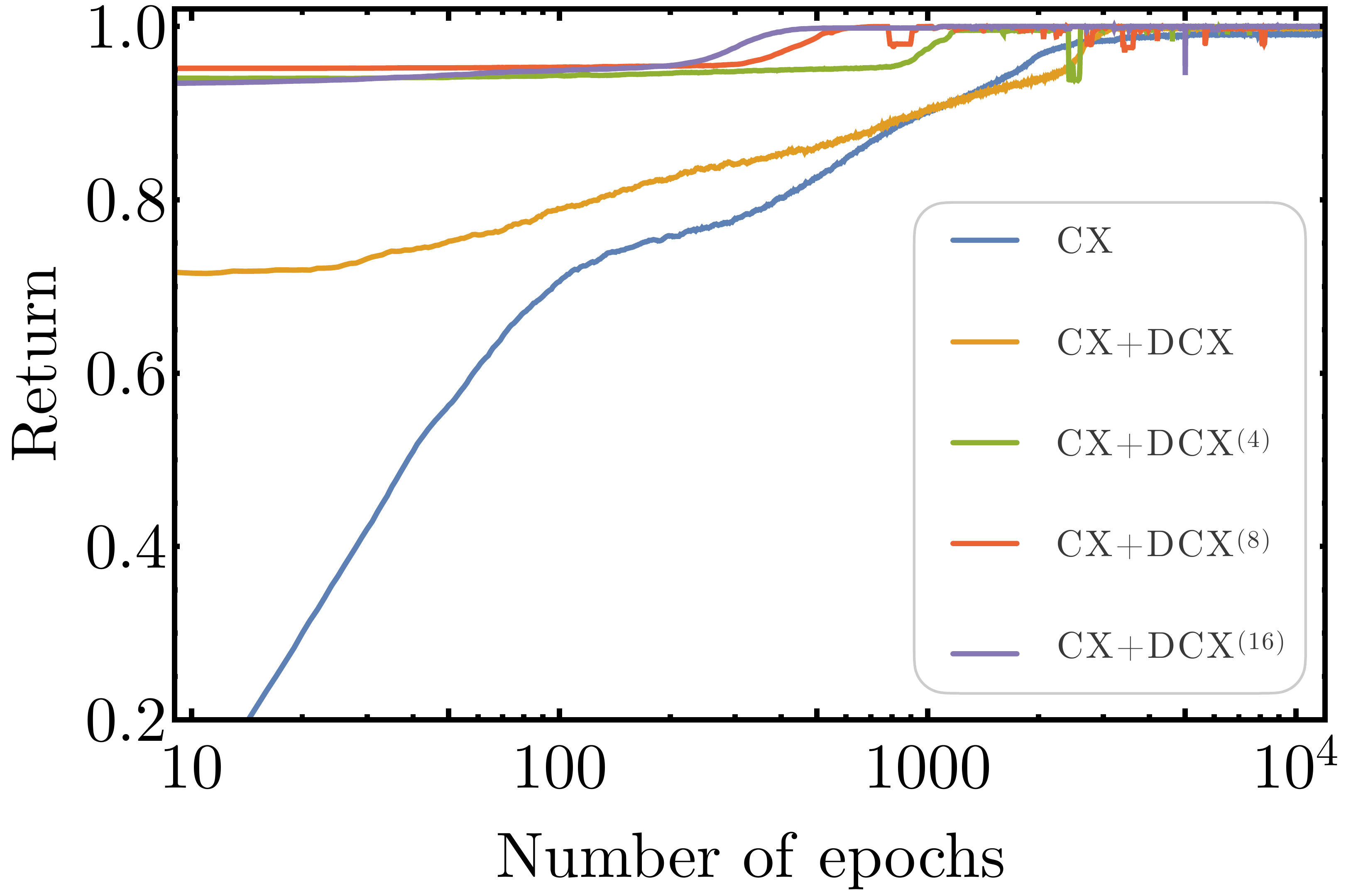}
    \caption{Speedup at finding encoders for $\code{21,1,5}$ codes by using gadgets. Results are averaged over 10 agents. Returns are normalized for easier comparison and the task is successful when the return is 1. More powerful gadgets provide much faster discovery.}
    \label{fig:speedup}
\end{figure}

\section{Results}
\label{sect:results}

\subsection{Facilitation of automated discovery of encoders with $ d = 5 $ 
            by RL with gadgets}
\label{sect:acceleration}

Medium-complexity encoders with $ d = 5 $ can be found with the help of the simple, i.e., CX-based, RL search. However, the gadget-based RL search using CX+DCX gadgets provides a substantial acceleration to discover solutions. As shown in Fig.\ref{fig:speedup} with the $\code{21,1,5}$ code as an example, the learning process exhibits markedly different convergence behaviors depending on the gadget configuration. The baseline CX-only approach requires approximately 6000 epochs to reach optimal return, displaying a gradual learning curve that starts from 0.2 and slowly increases. In contrast, the enhanced CX+DCX$^{(16)}$ configuration achieves near-optimal performance (return $\sim$1) in just 300 epochs, representing a 20x speedup in convergence time.

Interestingly, we observe that increasing the power of DCX gadgets (from DCX$^{(4)}$ to DCX$^{(8)}$ to DCX$^{(16)}$) leads to progressively faster convergence, with CX+DCX$^{(16)}$ showing the most rapid approach to discovery. The base CX+DCX configuration, while performing better than CX-only, demonstrates slower convergence compared to its counterparts with multiple DCX gadgets, suggesting that the multiplicity of DCX operations plays a crucial role in the search efficiency. Therefore, we believe that the power of the DCX gadget relies on it being useful for building more complex gadgets.

Another pronounced advantage of the gadget-based RL search is related to the substantially enhanced success rate. We define the success rate as the fraction of agents which are able to find solutions for a given set of hyperparameters. For the example shown in Fig.\ref{fig:speedup}, the success rate was approximately enhanced by a factor 2-3 when the powerful gadgets were used.

The significant advantage of the gadgets-based RL approach becomes even more pronounced when increasing  the encoder complexity. For example, the discovery of encoding circuits for [[31,7,5]] codes was accelerated by  $\times50$ if the DCX$^{(16)}$ gadget is used instead of only CX. Simultaneously, the success rate became 100\% (as opposed to $\sim 10\%$ produced by the CX-based search).

Finally, we note that both the speed of the search and the success rate depend on the choice of the hyperparameters. The results that we present in this paper have been obtained from our most successful attempts at fine-tuning hyperparameters and our discussion should be treated as a discussion of tendencies. As in any machine-learning scenario, the efficiency of the RL application can be further improved if one spends more 
computing time to further optimize hyperparameters.

\subsection{Discovery of large-distance codes, $ d \ge 6 $ at $ k = 1 $, by using gadgets}
\label{sect:n_d-diagr}

We have already mentioned that the CX-based RL search for encoders with $ k = 1 $ can be successful for distances $ d \le 5 $ and fails at $ d \ge 6 $. 
Using gadgets to explore $ d = 5 $ brings two obvious advantages: the search is crucially accelerated, and its success rate is noticeably enhanced, cf. Sect.\ref{sect:acceleration} and Fig.\ref{fig:speedup}. 

\begin{figure}[tb]
    \centering
    \includegraphics[width=0.95\linewidth]{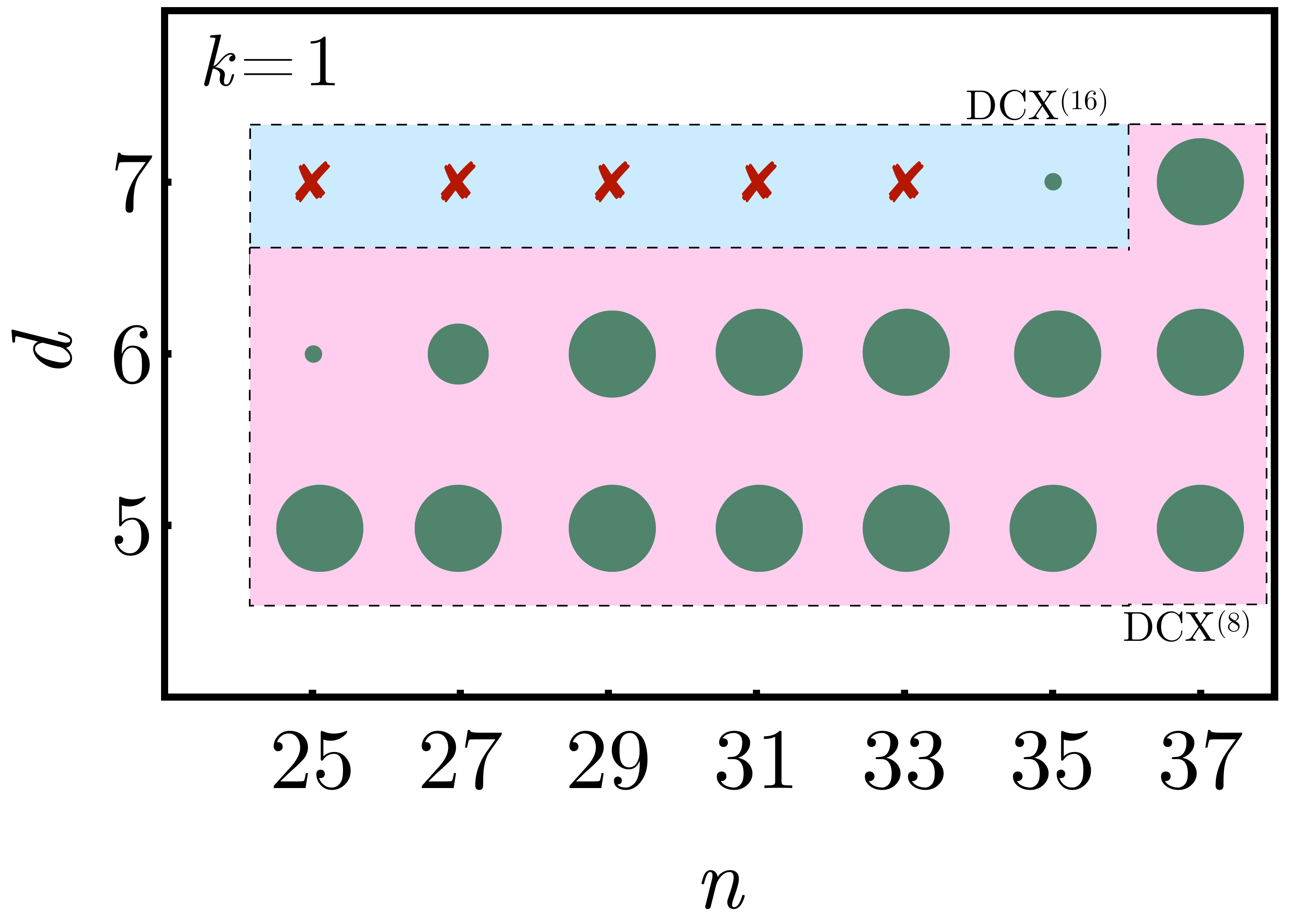}
    \caption{Efficiency of the gadget-based RL for the automated discovery of medium- to large distance encoders. The size of circles reflects the RL efficiency, i.e. the fraction of successful RL training runs, ranging from 100\% (large circles) to $<$10\% (the smallest circle). Colors of different regions mark the type of the most powerful gadgets that we used for the calculations. Red crosses denote runs which failed at a given n. Note that increasing n and/or employing more powerful gadgets can allow one to reach larger distances. Many codes of the type [[n,1,5]] can be found by using only the CX gates, though with substantially longer calculation times  and lower success rates.}
    \label{fig:n_d_k1}
\end{figure}

\begin{figure*}
    \centering
    \includegraphics[width=0.95\linewidth]{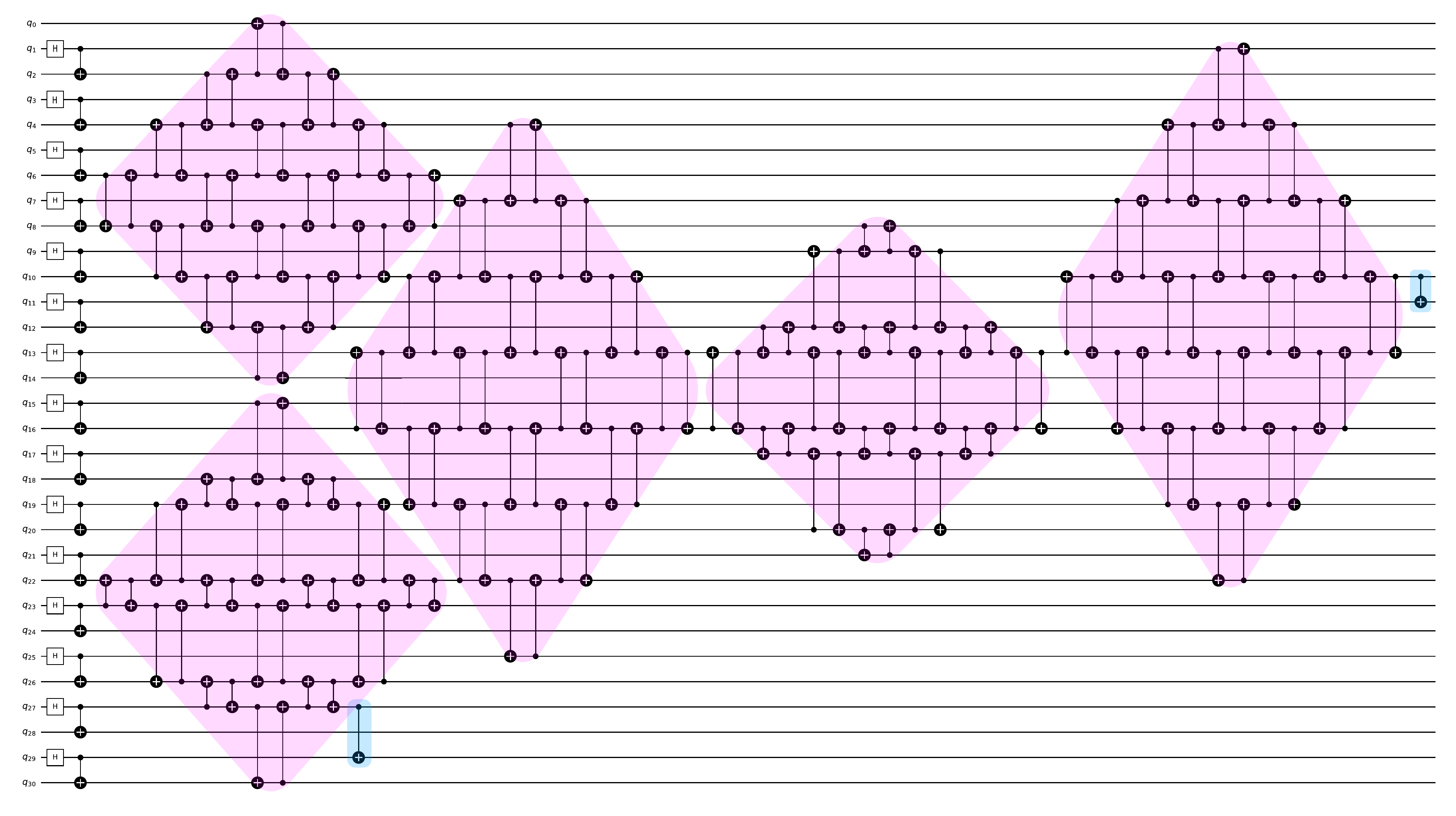}
    \caption{Example of $\code{31,1,6}$ code discovery with DCX$^{(8)}$ gadgets. In total, the agent uses 5 gadgets (highlighted in magenta), each consisting of 32 CX gates, and two single CX gates (highlighted in blue). The circuit is initialized with an equal number of Bell pairs. The total number of CX gates needed is 177 and the circuit depth is 55, which could be reduced to a depth of 29 by optimizing a pulse to implement a DCX gate on the hardware level as a single gate.}
    \label{fig:circuit_example_31_1_6}
\end{figure*}

The full power of the gadget-based RL search becomes obvious if one explores $ d = 6 $. There is a range of n values where novel (not known previously) encoders are discovered very quickly with a high success rate, see Fig.\ref{fig:n_d_k1}. Fig.\ref{fig:circuit_example_31_1_6} illustrates an example of such an encoder. We emphasize that we found the discovery of codes with $ d \ge 6 $ by the CX-based RL search to be impossible. 

The success rate of the gadget-based RL search becomes small if one focuses on the code [[25,1,6]] and fails at smaller values of $ n $. Hence, we observe the existence of a lower boundary for the RL application. We believe that it is directly related to a substantially decreasing number of successful solutions when $ n $ approaches its minimal theoretical value. For instance, the smallest $d=7$ CSS code is the Golay code $\code{23,1,7}$~\cite{Steane1996}.

The most striking result which we obtained by using the gadgets is the discovery of encoders with $ d = 7 $, in particular [[35,1,7]] and [[37,1,7]]. We did not attempt to reach higher distances but are confident that this task is feasible with a proper choice of the gadgets and, perhaps, with larger values of $ n $. The major limitation is likely to result from the amount of available GPU memory rather than from the approach itself. Moreover, these results were obtained with a single GPU, suggesting that a distributed approach on multiple GPUs can enhance the reachable code distance even more.

We would like to conclude this section by comparing the performance of our codes with that of surface codes. To reach the distance $ d $, the surface codes require $ n = d^2 $ qubits and, hence, have an encoding ratio $ n / d = d $. The majority of the codes with $ d = 6, 7 $ shown in Fig.\ref{fig:n_d_k1} have a smaller (or even substantially smaller) ratio $ n / d $ and, thus, have a higher encoding rate than surface codes. In other words, the high-distance encoding discovered by gadget-based RL can be implemented by using smaller qubit arrays and, hence, is less resource-consuming.

\subsection{Encoding many logical qubits: example of $ d = 6 $}
\label{sect:n_k-diagr}

\begin{figure}[tb]
    \centering
    \includegraphics[width=0.95\linewidth]{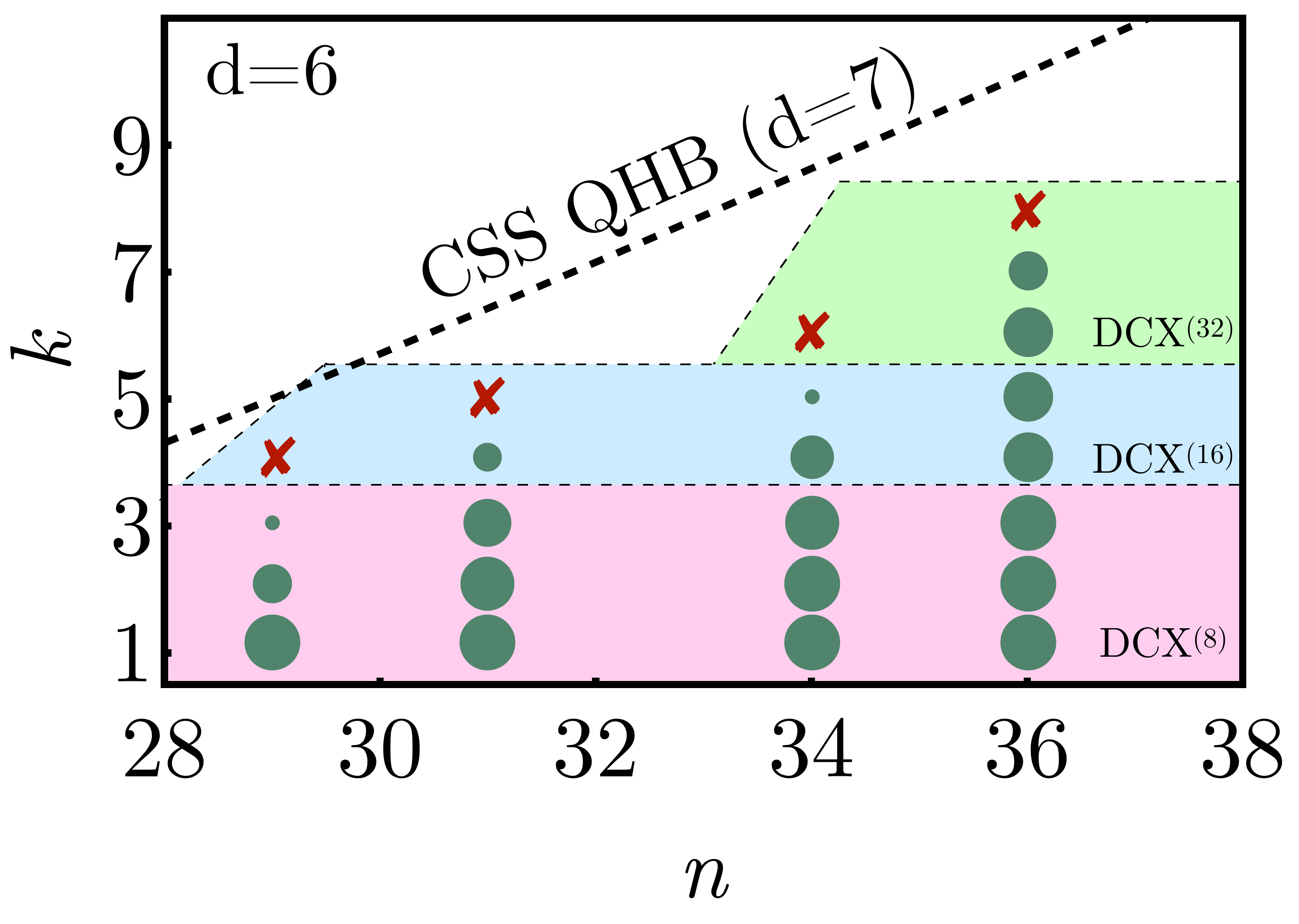}
    \caption{Efficiency of the gadget-based RL for the automated discovery of encoders for many logical qubits at $ d = 6 $. The meaning of circle sizes, colors of different layers and red crosses is explained in the caption of Fig.\ref{fig:n_d_k1}. Note that increasing n and/or employing more powerful gadgets allows one to reach a larger number of logical qubits. The dashed line is the Quantum Hamming Bound (QHB) for self-dual CSS codes (see Appendix \ref{appendix:QHB}). The QHB is a packing argument mapping each possible error pattern to a unique syndrome, and is only well-defined for odd code distances. Here, we show the bound for distance 7, meaning that codes above the dashed line are unlikely to exist, but below the line are very likely to exist.}
    \label{fig:n_k_d6}
\end{figure}

The hierarchy of gadgets, $\left\{ {\rm CX}, {\rm DCX}, {\rm DCX}^{(2q)}\right\}$ has been discovered in examples with $ k = 1 $. We could have repeated the gadget discovery process for simple encoders with $k > 1$, potentially leading to different gadgets being discovered. Instead, we have directly used the same already known gadgets for codes with $ k > 1 $. This approach was very useful and allowed us to discover the more challenging encoders ($ d > 5$) for encoding several logical qubits ($k > 1$).

Fig.\ref{fig:n_k_d6} shows the result of the gadget application for the discovery of codes with $ d = 6 $ and $ 1 \le k \le 7 $. We remind the reader that these codes are not reachable by the standard RL approach based only on using CX gates. The success rate is explained in the previous section. In general, the diagrams shown in Fig.\ref{fig:n_d_k1} and Fig.\ref{fig:n_k_d6} have several common features. In particular, the success rate at $ d = 6 $ and some fixed $ k $ always drops with decreasing $ n $. On the other hand, by increasing $ n $ and using more powerful gadgets, we were able to find the encoders for relatively large values of $ k $. The best achievement for our choice of hyperparameters is the code $ [[36, 7, 6]] $. We would like to emphasize that this is not the ultimate limit of our approach. We are confident that one can extend our results to even larger values of $ k $ after fine-tuning the hyperparameters and, perhaps, using more complicated gadgets at larger $ n $. In particular, as shown in Fig.\ref{fig:n_k_d6} (dashed line), the code parameters of the discovered encoding circuits lay below the Quantum Hamming Bound (QHB) for CSS codes; see Appendix \ref{appendix:QHB} for details on this bound. The main limitation here is related to the availability of GPU memory.

Based on the successful use of the gadgets $\left\{ {\rm CX}, {\rm DCX}, {\rm DCX}^{(2q)}\right\}$ for codes with $ k > 1 $ we can conclude that this hierarchy of gadgets possesses some universality. The origin of this universality and the possible existence of other families of powerful gadgets remain open questions which we postpone for future studies.

Finally, let us emphasize that the codes which have been discovered by the gadget-based RL approach may have a promising encoding efficiency reflected by the ratio $ k / n $. For example, the code $ [[36, 7, 6]] $ has $ k/n = 7/36 $ which is slightly better than the encoding efficiency of the LDPC code [[72,12,6]] of \cite{bravyi2023highthreshold}, namely $ k/n = 1/6 $. However, the price paid for this higher encoding rate is having generators with larger weights, see the discussion in Sect.\ref{sect:weights}.

\subsection{Weights of generators}
\label{sect:weights}

\begin{figure}
    \centering
    \includegraphics[width=0.95\linewidth]{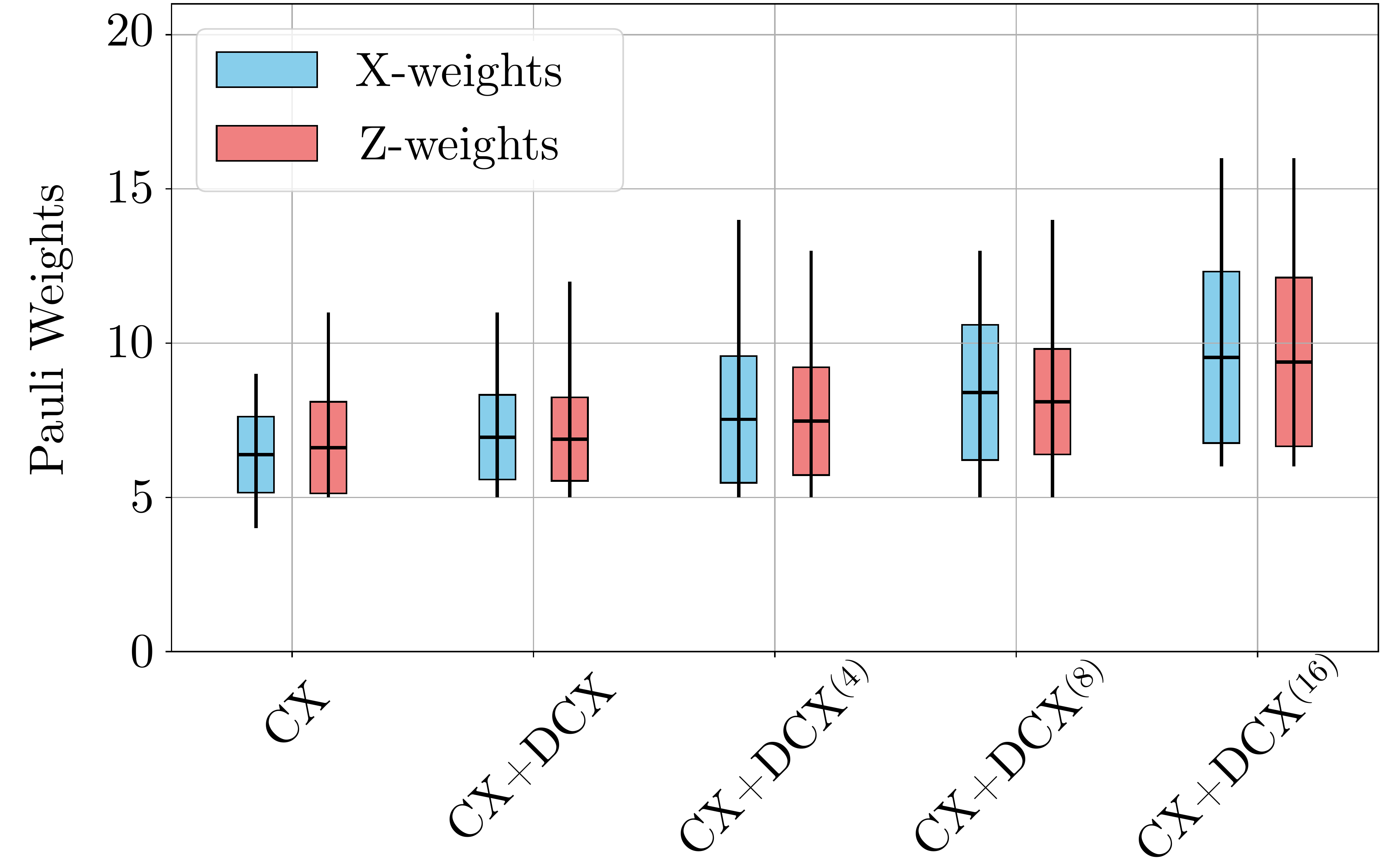}
    \caption{Weights of $\code{21,1,5}$ codes obtained by using different gadgets. Results are averaged over 10 agents. The body of the candle shows the mean$\pm$standard deviation, and the vertical black lines go from minimum to maximum. More powerful gadgets lead to codes with larger weights on average.}
    \label{fig:weights_21_1_5}
\end{figure}

\begin{figure}
    \centering
    \includegraphics[width=0.95\linewidth]{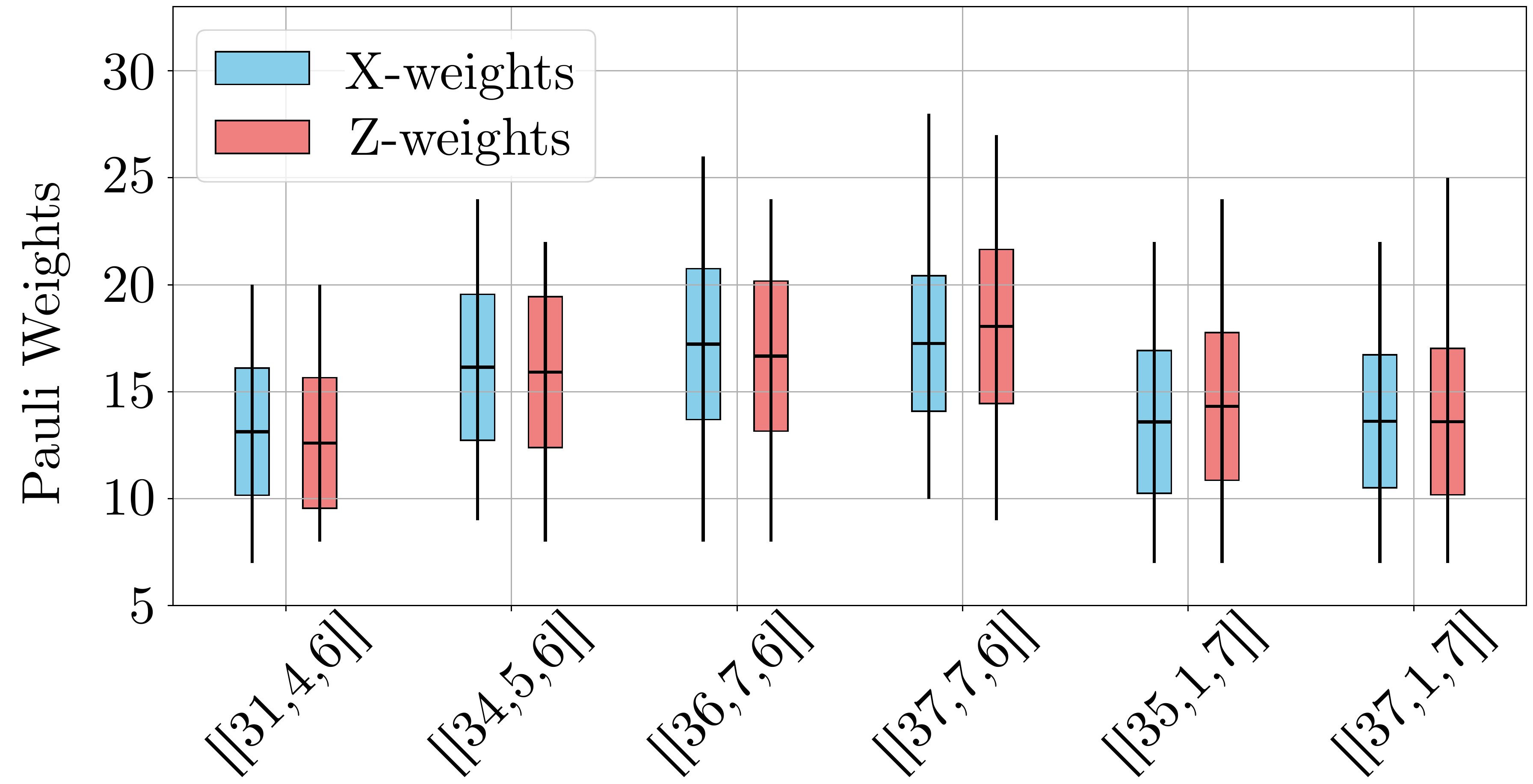}
    \caption{Weights of the largest codes that we have been able to find. Results are averaged over successful runs. Notations are the same as in Fig.\ref{fig:weights_21_1_5}. The gadgets employed have been (from left to right): DCX$^{(16)}$, DCX$^{(16)}$, DCX$^{(32)}$, DCX$^{(32)}$, DCX$^{(16)}$ and DCX$^{(8)}$.}
    \label{fig:weights_d6_d7}
\end{figure}

We observed one seemingly general limitation of our gadget-based approach, when applied to this QEC task: the relatively high weight of the discovered code generators. We include in Fig.~\ref{fig:weights_21_1_5} a comparison of the weights of the discovered $\code{21,1,5}$ codes with different levels of gadget complexity. There we see that, on average, weights of codes found with DCX$^{(16)}$ are 50 $\%$ larger than those found by using CX only. However, never do we encounter a situation where the average weight is larger than $n/2$. Weights can be as small as 4 and as large as 16, depending on the gadgets used.

The (generally undesirable) generation of large weights can be formally understood from the transformation rules of the DCX$^{(2m)}$ gadgets. In particular, a code generator's weight can go from 1 to $m$ after a single gadget operation, see Appendix \ref{appendix:analytics_gadgets} for an analytic treatment.

We also examine the larger codes with $d=6$ and $d=7$ that we have presented in Sections \ref{sect:n_d-diagr} and \ref{sect:n_k-diagr} and show a representative analysis in Fig.~\ref{fig:weights_d6_d7}. The codes shown were selected as the ones with the most efficient encoding rates. The majority of the weights fall within the range of 10-20, with a few outliers both below 10 and above 25. We also don't see significant jumps in the resulting weights when increasing the complexity of the gadget used. For instance, codes $\code{34,5,6}$ and $\code{36,7,6}$ were found with DCX$^{(16)}$ and DCX$^{(32)}$ gadgets, respectively, yet their weights are rather similar, 
see Fig.\ref{fig:weights_d6_d7}. Another interesting feature seen in Fig.~\ref{fig:weights_d6_d7} is that the $d=7$ codes possess smaller weights on average than the other $d=6$ codes shown. However, for comparison, the smallest $d=7$ code is the Golay code, with parameters $\code{23,1,7}$ and generators of weight 8~\cite{Steane1996}. The alternative using surface codes is the $\code{49,1,7}$ code with generators of weight 4 or 2.

While modifying the reward function to penalize high-weight generators provides some improvement, the reduction remains modest. In our experiments, adding a weight penalty term to the reward function yielded only a $10-15\%$ weight reduction, but the success rate of agents was hampered. We note that slightly lower weights arose automatically when only employing CNOT gates without any deliberate modification of the reward to assign a preference to low weights, see Fig.\ref{fig:weights_21_1_5}. However, despite our various optimization attempts, we have not yet matched the low-weight efficiency of surface codes. This limitation stems partly from our dual optimization goal - we seek not only minimal-weight generators but also efficient implementation circuits for specific hardware constraints. This combined objective appears to make weight optimization particularly challenging. Overall, this suggests that achieving low-weight generators of the RL-discovered codes while maintaining hardware-efficient circuits may require fundamental changes to the approach rather than simple reward modifications. For instance, a recent interesting approach that discovers codes (without hardware-efficient encoders) of low weight is presented in Ref.\cite{he2025discoveringhighlyefficientlowweight}.

\section{Discussion and conclusions}
\label{sect:discussion}

We have demonstrated that the reinforcement learning approach
to the discovery of quantum circuits can be substantially
facilitated by using composite gates – the gadgets. Our approach
consists of several basic steps. First, one generates simple
circuits by using the vanilla RL method  \cite{olle_2024}. Next, the available circuits
are preprocessed in order to enable a posterior tractable visual inspection. At this stage, our algorithm filters out the equivalent
circuits with the same canonical tableaux and rearranges qubits 
and gates in the remaining non-equivalent circuits to compress the dataset into a handful of representative circuits.
Finally, one can use the first few discovered gadgets to generalize to larger qubit numbers. 

We have applied this new approach to the example of discovering
codes for quantum error correction. In this application, we were able to conjecture a hierarchy of gadgets involving two, four,
eight, etc., qubits. The use of gadgets from this family has
accelerated the RL search of new QEC codes by one to two orders of magnitude. 
This made it possible to scale the RL-based automatic
discovery of codes to larger numbers of data qubits and larger code
distances than without gadgets. Using the standard $ [[n, k, d]] $ notation, we have
found encoders for codes with $ k = 1, d \le 7 $
and $ d = 6, k \le 7 $. This range goes well beyond the results
of Ref.\cite{olle_2024} where the RL method with primitive 
Clifford gates was used. To the best of our knowledge, the
encoders, e.g., for the codes $ [[n,1,7]] $ and $ [n,7,6]] $ are
reported in the current paper for the first time. 

Two important notes are due here. Firstly, the total number 
of qubits, $ n $, can be fine-tuned to improve the performance 
of the automated discovery process of codes and, simultaneously, 
competitive ratios $ d / n $ and $ k / n $ can be reached. The ratios that we have managed to achieve can be better than those of the well-known surface- and LDPC codes~\cite{bravyi2023highthreshold}. Secondly, the principal computational limitation 
of our approach comes from the amount of the available computer (GPU) memory.

The main disadvantage of the encoders found with the help of the
gadget-based RL is the rather large weights of code generators. This seems to
be an unavoidable consequence of the use of gadgets in the one-dimensional networks of qubits. 
We do not exclude that the weights are reduced if the
gadgets are used in different physical networks of qubits, for example, in two-dimensional qubit structures which mimic the 
connectivity of surface codes and which match the current capabilities of available hardware better. In this scenario, it would also be interesting to leverage symmetries to restrict the search space further by either imposing them in the implementation of actions or by employing tools from geometric deep learning.

Complicated gadget-based encoders contain a large number of primitive Clifford gates. 
Current quantum hardware has been designed by optimizing each primitive gate separately. However, there is no need to control individually those primitives which are part of intensively used gadgets. Therefore, a higher-level optimization might be desirable. In particular,
one may try to engineer hardware where each gadget
works as an irreducible unit and, hence, is controlled by a single control pulse, which is specially-designed for a given platform. This may substantially improve the performance and operation speed of the encoders.

We do not know whether the gadget family that we have found is
unique. A possibility would be that other families could also be successfully used to facilitate
the RL discovery of new quantum circuits. The search for alternative
gadgets is currently based on the visual analysis of the circuits, which may become difficult in other scenarios.
Therefore, it is highly desirable to automate this step. One possibility could be that the
preprocessed circuits be represented as two-dimensional
graphs and further analyzed with the help of standard graph-based algorithms.
We will address the automated search of gadgets elsewhere.

Overall, our approach paves the way for scaling the RL-based 
automatic discovery of more complicated quantum circuits. While we have presented the particular case of  QEC encoders, we believe that the same gadget-based approach can also be used in other tasks, such as, e.g.,  in the automated discovery of logical operations between logical qubits or 
of quantum algorithms.

\acknowledgements

We acknowledge enlightening and helpful discussions with Remmy Zen, Maximilian N{\"a}gele, and Matteo Puviani,  as well as
Christian Schilling and members of his group. This research is part of the 
Munich Quantum Valley network, which is supported by the Bavarian state government with funds
from the Hightech Agenda Bayern Plus.

\bibliography{refs.bib}

\newpage
\appendix

\section{The Quantum Hamming Bound for CSS codes}
\label{appendix:QHB}

The quantum Hamming bound (QHB) establishes a theoretical limit on QEC codes. For an [[n,k,d]] stabilizer code, the bound is
\begin{equation}
    2^{n-k} \geq \sum_{j=0}^{t} 3^j \binom{n}{j}~,  \label{eq:QHB_stabilizer}
\end{equation}
where $t = \lfloor(d-1)/2 \rfloor$ is the maximum weight of the errors that the code can correct. The summation term in Eq.\eqref{eq:QHB_stabilizer} represents the number of possible error patterns up to weight t. Assuming that each error gets mapped to a different syndrome gives the upper bound in terms of $n$ and $k$.

The factor $3^j$ appears because errors can be of three types: bit flips (X errors), phase flips (Z errors), and combined bit-phase flips (Y errors).

In CSS codes, X errors and Z errors are detected independently. Assuming a scenario in which we want to correct an equal number of errors of both X and Z-type, the optimal configuration is to have an equal number of X  and Z code generators, equal to $\lfloor(n-k)/2 \rfloor$. This leads to the notion of weakly self-dual CSS codes. Thus, we can write a version of the QHB bound for weakly self-dual codes as
\begin{equation}
    2^{\lfloor(n-k)/2 \rfloor} \geq \sum_{j=0}^{t} \binom{n}{j}~.
\end{equation}
Notice that this bound offers a necessary but not sufficient condition for a CSS code with parameters $[[n,k,d]]$ to exist. Moreover, this assumes that the code is non-degenerate (in the sense of different errors having different syndromes). In particular, being in a regime where this bound is satisfied does not guarantee existence of a code, but it does exclude the existence of non-degenerate codes when it is violated. 

As a final remark, codes that saturate this bound are called \textit{perfect}, and an example is the $[[23,1,7]]$ self-dual CSS code, as can be easily verified.

\section{Transformation Rules of Gadgets}
\label{appendix:analytics_gadgets}

The CNOT transformation rules are the following:
\begin{align}
\Qcircuit @C=1em @R=.7em {
    & \ctrl{1} & \qw & \nghost{\mathtt{XI} \to \mathtt{XX}~,~\mathtt{ZI} \to \mathtt{ZI}} & \lstick{\mathtt{XI} \to \mathtt{XX}~,~\mathtt{ZI} \to \mathtt{ZI}}\\
    & \targ & \qw & \nghost{\mathtt{IX} \to \mathtt{IX}~,~\mathtt{IZ} \to \mathtt{ZZ}} & \lstick{\mathtt{IX} \to \mathtt{IX}~,~\mathtt{IZ} \to \mathtt{ZZ}}
}    \label{eq:cx_ct_rule}
\end{align}

\begin{align}
\Qcircuit @C=1em @R=.7em {
    & \targ & \qw & \nghost{\mathtt{XI} \to \mathtt{XI}~,~\mathtt{ZI} \to \mathtt{ZZ}} & \lstick{\mathtt{XI} \to \mathtt{XI}~,~\mathtt{ZI} \to \mathtt{ZZ}}\\
    & \ctrl{-1} & \qw & \nghost{\mathtt{IX} \to \mathtt{XX}~,~\mathtt{IZ} \to \mathtt{IZ}} & \lstick{\mathtt{IX} \to \mathtt{XX}~,~\mathtt{IZ} \to \mathtt{IZ}}
}    \label{eq:cx_tc_rule}
\end{align}
We note that we can obtain the transformation rules for $Z$ operators by exchanging control with target and $Z$ with $X$. This property is also true for all other gadgets that we consider in this paper. Thus, from now on, we will ignore the second orientation and study the transformation rules of more complex gadgets in the first orientation (see Fig.~\ref{fig:gadgets}).

The transformation rules for DCX are:
\begin{align}
\Qcircuit @C=1em @R=.7em {
    & \ctrl{1} & \targ & \qw &\nghost{\mathtt{XI} \to \mathtt{IX}~,~\mathtt{ZI} \to \mathtt{ZZ}} & \lstick{\mathtt{XI} \to \mathtt{IX}~,~\mathtt{ZI} \to \mathtt{ZZ}}\\
    & \targ & \ctrl{-1} & \qw &\nghost{\mathtt{IX} \to \mathtt{XX}~,~\mathtt{IZ} \to \mathtt{ZI}} & \lstick{\mathtt{IX} \to \mathtt{XX}~,~\mathtt{IZ} \to \mathtt{ZI}}
}    \label{eq:dcx_ct_rule}
\end{align}
The transformation rules for DCX$^{(4)}$ are:
\begin{align}
    &  \mathtt{XIII} \to \mathtt{XIXI}~, & \mathtt{ZIII} \to \mathtt{IZZI} \label{eq:dcx4_rule_1} \\
    &  \mathtt{IXII} \to \mathtt{IXXX}~, & \mathtt{IZII} \to \mathtt{ZZZZ} \label{eq:dcx4_rule_2} \\
    &  \mathtt{IIXI} \to \mathtt{XXXX}~, & \mathtt{IIZI} \to \mathtt{ZZZI} \label{eq:dcx4_rule_3} \\
    &  \mathtt{IIIX} \to \mathtt{IXXI}~, & \mathtt{IIIZ} \to \mathtt{IZIZ} \label{eq:dcx4_rule_4}
\end{align}
We also see another symmetry between the $X$ and $Z$ rules: rule \eqref{eq:dcx4_rule_4} is identical to rule \eqref{eq:dcx4_rule_1} after inverting left and right and exchanging X by Z. The same happens for rules \eqref{eq:dcx4_rule_2} and \eqref{eq:dcx4_rule_3}. Thus, we only need to write the $X$ rules moving forward.
These are the transformation rules for DCX$^{(8)}$:
\begin{align*}
    & \mathtt{XIIIIIII} \to \mathtt{XIXIXIII}~,\\
    & \mathtt{IXIIIIII} \to \mathtt{IXXXIXII}~,\\
    & \mathtt{IIXIIIII} \to \mathtt{XXIIXIXI}~,\\
    & \mathtt{IIIXIIII} \to \mathtt{IXIIXXIX}~,\\
    & \mathtt{IIIIXIII} \to \mathtt{XIXXXIIX}~,\\
    & \mathtt{IIIIIXII} \to \mathtt{IXIXIXXI}~,\\
    & \mathtt{IIIIIIXI} \to \mathtt{IIXIIXXX}~,\\
    & \mathtt{IIIIIIIX} \to \mathtt{IIIXXIXI}~,
\end{align*}
where we see that the maximum weight is 5. 

\begin{figure}
    \centering
    \includegraphics[width=0.95\linewidth]{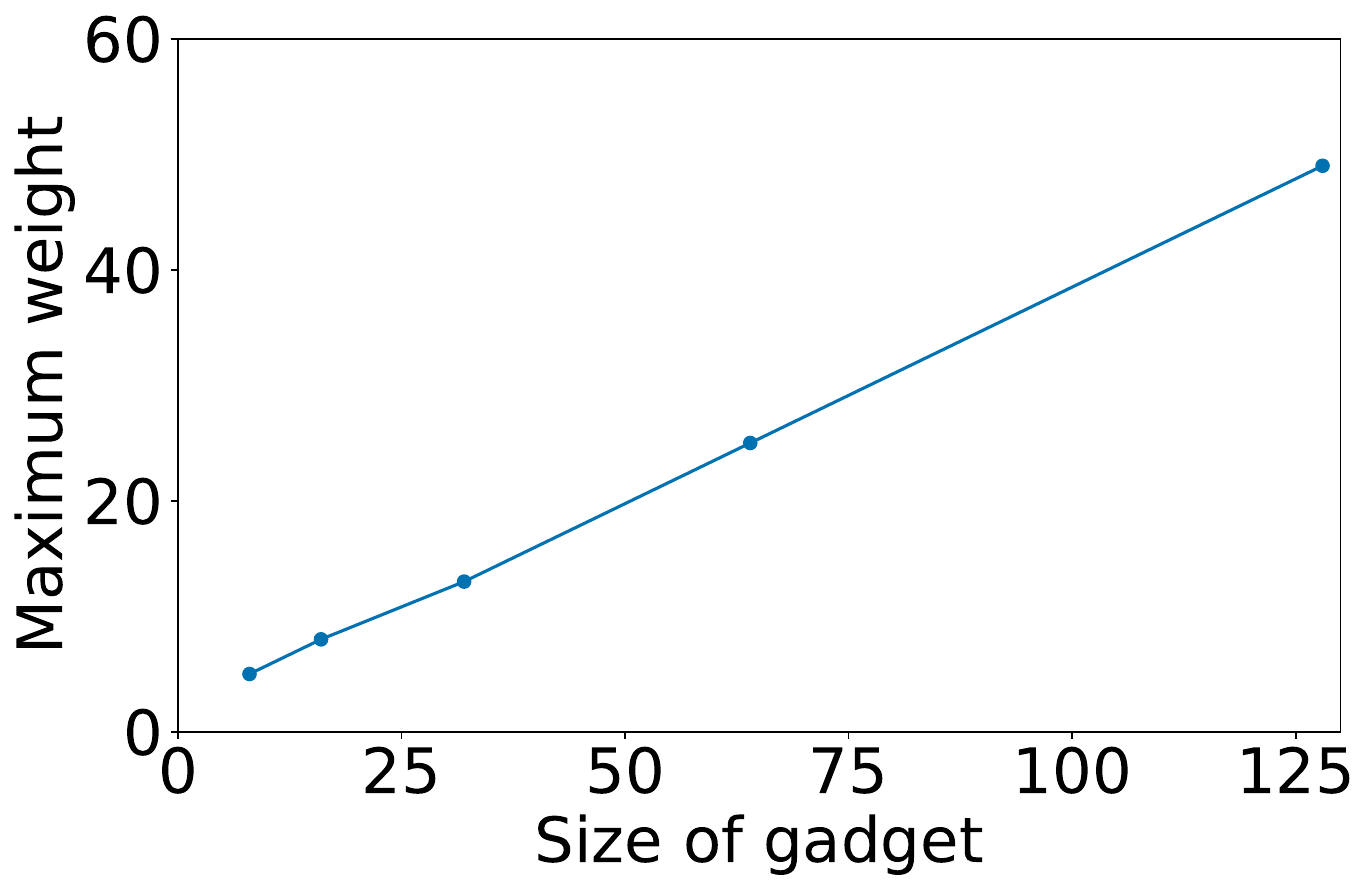}
    \caption{Maximum weight of weight-1 Paulis after being propagated through a gadget of size $m$, DCX$^{(m)}$. The maximal weight is close to but below $m/2$.}
    \label{fig:gadget_weights}
\end{figure}

Similar rules can be derived for higher-order gadgets like DCX$^{(16)}$, etc. Since they are much more lengthy and not particularly illuminating, we refrain from writing them down explicitly, but we show their trend in Fig.\ref{fig:gadget_weights}.

Hence, we observe that the maximum weight gradually increases with increasing the maximal order $ m $ of the utilized gadget, DCX$^{(m)}$. 
Moreover, the maximal weight is close to $m/2$, see Fig.\ref{fig:gadget_weights}.  This might explain the growth of the weights discussed in Sect.\ref{sect:weights}. 

\end{document}